\documentclass{article}
\usepackage[centertags]{amsmath}
\DeclareMathOperator{\SN}{sn} \DeclareMathOperator{\CN}{cn}
\DeclareMathOperator{\FF}{F} \DeclareMathOperator{\EE}{E}
\begin{document}

\title{Direct integral method, complete discrimination system for polynomial and applications to classifications
of all single traveling wave solutions to nonlinear differential
equations:a survey}

\author{ Chengshi Liu \\Department of Mathematics\\Daqing Petroleum Institute\\Daqing 163318, China
\\Email: chengshiliu-68@126.com\\csliu@dqpi.edu.cn}

 \maketitle

\begin{abstract}
Complete discrimination system for polynomial and direct integral
method were discussed systematically. In particularly, we pointed
out some mistaken viewpoints. Combining with complete discrimination
system for polynomial, direct integral method was developed to
become a powerful method and was applied to a lot of nonlinear
mathematical physics equations. All single traveling wave solutions
to theses equations can be obtained. As examples, we gave all
traveling wave solutions to some equations such as mKdV equation,
Sine-Gordon equation ,Double Sine-Gordon equation, triple
Sine-Gordon equation, Fujimoto-Watanabe equation, coupled Harry-Dym
equation and coupled KdV equation and so on.
At the end, we give the partial answers to Fan's a problem.\\


\end{abstract}

\section{Introduction}
In order to solve the traveling wave solutions to some nonlinear
evolution equations, sometimes these equations are reduced to an
ordinary differential equations such as
\begin{equation}
u'(\xi)=G(u,\theta_1,\cdots,\theta_m),
\end{equation}
where $ \theta_1,\cdots,\theta_m $ are parameters,its solutions
can be wrote by integral form
\begin{equation}
\xi-\xi_0=\int \frac{\mathrm{d}u}{G(u,\theta_1,\cdots,\theta_m)}.
\end{equation}
Thus what we  need is only to solve the integral (2).  According
to different parameters, we will give different solution to the
integral. This is so called direct integral method. Although
direct integral method is a routine method, but to decide the
parameter's scope is still rather difficult. Thus the most
important steps are to decide the parameter's scopes and the
corresponding integral solutions. A new mathematical tool named
complete discrimination system for polynomial([1-5]) is applied to
this problem so that the parameter's scopes is solved.\\

It is just because of combining with complete discrimination system
that direct integral method becomes a powerful and efficient method.
There are a lot of nonlinear mathematical physics equations such as
KdV equation, Hirota equation, breaking soliton equation, Coupled
mKdV equation,Chen-lee-Liu equation, Kundu equation equation,cubic
Schr\"{o}dinger equation, Sine-Gordon equation, double Sine-Gordon
equation, Sinh-Gordon equation, double Sinh-Gordon equation, triple
Sinh-Gordon equation, Combine KdV-mKdV equation and nonlinear
Klein-Gordon equation and so on, their all traveling wave solutions
can be obtained by using direct integral method. These complete
results can't be obtained by other any indirect methods ( see
[6-19]) such as trial equation method([19]), Jacobi elliptic
function expanse method ([12]), sub-equation method([15,16] and
homogenous balance method and so on.

Although direct integral method is the most basic method, but
there are two obvious mathematical facts about the integral(2) to
be ignored. One fact is that different variable transformation
don't offer to a different solution, different variable
transformation only give different representation of solution.
Another fact is that it is just different parameter's scope to
offer to different solution. Some authors didn't notice the fact
one([20-23]), they still think that a new variable transformation
can bring new solutions. We must point out that this is a mistaken
viewpoint. In fact, the elementary calculus tell us that a defined
function has infinite primitive functions which  are different
each other by a constant, and that a new variable transformation
only offer to a new representation of the primitive function.

In the present paper, using complete discrimination system for
polynomial and direct integral method ,we give the classifications
of all single traveling wave solutions to a lot of  nonlinear
differential equations such as mKdV equation and Sine-Gordon
equation, Fujimoto-Watanabe equation, coupled Harry-Dym equation and
coupled KdV equation and so on. An the
end, we give the partial answer to Fan's a problem.\\

This paper is organized as follows. In section 2,  we deal with some
equations whose reduced ODE's take the form of
$(u')^2=a_3u^3+a_2u^2+a_1u+a_0$, and  obtain the classifications of
all single traveling wave solutions to those equations. In section
3, we deal with some equations whose reduced ODE's take the form of
$(u')^2=a_4u^4+a_3u^3+a_2u^2+a_1u+a_0$, and  obtain the
classifications of all single traveling wave solutions to those
equations. In section 4, we give some applications of the complete
discrimination system for the fifth order polynomial.  In section 5,
we deal with two equations with rational form of reduced equations
and obtain classifications of their all single traveling wave
solutions. In section 6, we use a one-parameter's thick obtain the
classifications of all single traveling wave solutions to some
equations such as coupled KdV equation and so on. In section 7, we
discuss Fan's two problems, and give an answer to the first problem.

\section{Applications of the complete discrimination system for the third order polynomial}

There are a lot of nonlinear evolution equations such as  KdV
equation
\begin{equation}
u_t+6uu_x+u_{xxx}=0;
\end{equation}
BBM equation
\begin{equation}
u_t+u_x+uu_x+\alpha u_{xxt}=0;
\end{equation}
Klein-Gordon equation
\begin{equation}
u_{tt}-u_{xx}+\alpha u+\beta u^3=0;
\end{equation}
Landau-Ginburg-Higgs eqution
\begin{equation}
u_{tt}-u_{xx}-m^2 u+g^2u^3=0;
\end{equation}
Sine-Gordon eqution
\begin{equation}
u_{xt}=a\sin u;
\end{equation}
Dodd-Bullough-Mikhailov equation
\begin{equation}
u_{xt}+p\exp(u)+q\exp(-2u)=0;
\end{equation}
Zhiber-Shabaut equation
\begin{equation}
u_{xt}+p\exp(u)+q\exp(-u)+r\exp(-2u)=0;
\end{equation}
Sinh-Gordon eqution
\begin{equation}
u_{xt}=a\sinh u;
\end{equation}
Joseph-Egri equation Sine-Gordon eqution
\begin{equation}
u_t+u_x+uu_x+\beta u_{xtt}=0;
\end{equation}
Kundu equation
\begin{equation}
\mathrm{i}u_t-u_{xx}-2\mathrm{i}(2\alpha-1)|u|^2u_x*+\alpha(4\alpha-1)|u|^4u=0;
\end{equation}
derivation Shrodinger eqution
\begin{equation}
u_t=\mathrm{i}u_{xx}+\mathrm{i}(c_3|u|^2+c_5|u|^4)u+s_2(|u|^2u)_x=0;
\end{equation}
Ablowitz equation
\begin{equation}
\mathrm{i}u_t=u_{xx}-4\mathrm{i}(|u|^2u_x*+8|u|^4u;
\end{equation}
and so on. They  all can be reduced to the following third order
integrable ODE
\begin{equation}\label{3}
(u')^2=a_3u^3+a_2u^2+a_1u+a_0,
\end{equation}
So if we give the classification of all solutions of the Eq.(15),
then all single traveling wave solutions to those equations (3)-(14)
can be classified. According to the Ref.\cite{Liu}, we list those
results for convenience in the following. let
\begin{equation}
w=(a_3)^{\frac{1}{3}}u, \ \ d_2=a_2(a_3)^{-\frac{2}{3}}, \ \
d_1=a_1(a_3)^{-\frac{1}{3}}, \ \ d_0=a_0.
\end{equation}\\
Then Eq.(\ref{3}) becomes
\begin{equation}
 \pm(a_3)^{\frac{1}{3}}(\xi-\xi_0)=\int\frac{1}{\sqrt{w^3+d_2w^2+d_1w+d_0}}\,\mathrm{d}w.
\end{equation}\\
Denote
\begin{equation}
F(w)=w^3+d_2w^2+d_1w+d_0.
\end{equation}
\begin{equation}
\triangle=-27(\frac{2d_2^3}{27}+d_0-\frac{d_1d_2}{3})^2-4(d_1-\frac{d_2^2}{3})^3,
\end{equation}
\begin{equation}
D_1=d_1-\frac{d_2^2}{3},
\end{equation}\\
where $ \triangle $ and $ D_1 $ make up a complete discrimination
system for $ F(w) $.
There are the following four cases to be discussed(\cite{Liu}):\\

Case 1: $ \triangle=0, D_1<0 $, then we have $
F(w)=(w-\alpha)^2(w-\beta), \alpha\neq\beta. $ If $ w>\beta $, the
solutions are as follows:
\begin{equation}
u=(a_3)^{-\frac{1}{3}}[(\alpha-\beta)\tanh^2(\frac{\sqrt{\alpha-\beta}}{2}(a_3)^{\frac{1}{3}}(\xi-{\xi}_0))+\beta],\
\ \alpha>\beta;
\end{equation}
\begin{equation}
u=(a_3)^{-\frac{1}{3}}[(\alpha-\beta)\coth^2(\frac{\sqrt{\alpha-\beta}}{2}(a_3)^{\frac{1}{3}}(\xi-{\xi}_0))+\beta],\
\ \alpha>\beta;
\end{equation}
\begin{equation}
u=(a_3)^{-\frac{1}{3}}[(-\alpha+\beta)\sec^2(\frac{\sqrt{-\alpha+\beta}}{2}(a_3)^{\frac{1}{3}}(\xi-{\xi}_0))+\beta],
\ \alpha<\beta.
\end{equation}\\

Case 2: $ \triangle=0, \ D_1=0 $, then we have $ F(w)=(w-\alpha)^3
$, the solution is as follows:
\begin{equation}
u=4(a_3)^{-\frac{2}{3}}(\xi-{\xi}_0)^{-2}+\alpha.
\end{equation}\\

Case 3: $ \triangle>0, \ D_1<0 $, then $
F(w)=(w-\alpha)(w-\beta)(w-\gamma) $, we suppose that $
\alpha<\beta<\gamma $.  When $ \alpha<w<\beta $, we have
\begin{equation}
u=(a_3)^{-\frac{1}{3}}[\alpha+(\beta-\alpha)\SN^2(\frac{\sqrt{\gamma-\alpha}}{2}(a_3)^{\frac{1}{3}}(\xi-{\xi}_0)),m)].
\end{equation}\\
When $ w>\gamma $, we have
\begin{equation}
u=(a_3)^{-\frac{1}{3}}[\frac{\gamma-\beta\SN^2(\frac{\sqrt{\gamma-\alpha}}{2}(a_3)^{\frac{1}{3}}(\xi-{\xi}_0)),m)}{\CN^2(\frac{\sqrt{\gamma-\alpha}}{2}(a_3)^{\frac{1}{3}}(\xi-{\xi}_0)),m)}],
\end{equation}\\
where $ m^2=\frac{\beta-\alpha}{\gamma-\alpha} $.\\[3mm]

Case 4: $ \triangle<0 $, then we have $ F(w)=(w-\alpha)(w^2+pw+q),
\ \ p^2-4q<0. $  We have
\begin{equation}
u=(a_3)^{-\frac{1}{3}}[\alpha-\sqrt{\alpha^2+p\alpha+q}+\frac{2\sqrt{\alpha^2+p\alpha+q}}{1+\CN((\alpha^2+p\alpha+q)^{\frac{1}{4}}(a_3)^{\frac{1}{3}}(\xi-{\xi}_0),m)}],
\end{equation}\\
where $ m^2=\frac{1}{2}(1-\frac{\alpha+\frac{p}{2}}{\sqrt{\alpha^2+p\alpha+q}}) $.\\

We take only three equations as examples to illustrate this
method, other equations can be dealt with similarly.\\

\textbf{Example 1}.Classification of all single traveling wave
solutions
to KdV equation\\

Although KdV equation is a classical equation, its single traveling
wave solutions had been discussed many years ago. But its
classification hasn't been reported, some author also try to give
its new single traveling wave solutions, so we take it as an example
and give its all single traveling wave solutions. In fact under the
traveling wave transformation, KdV equation can be reduced to the
following ODE:
\begin{equation}\label{KDV}
(u')^2=-\frac{2}{k^2}u^3-\frac{\omega}{k^3}u^2+c_1u+c_0.
\end{equation}
Thus it is easy to give the classification of all solutions to ODE
\ref{KDV} according to the former results. For briefly we omit
them.\\

\textbf{Example 2}.Classification of all single traveling wave
solutions to Sine-Gordon
equation.\\

Sine-Gordon equation reads
\begin{equation}\label{SG}
u_{xt}=a\sin u,
\end{equation}
which widely applied in physics and engineering. Its so-called new
traveling wave solutions have been obtained in different functional
forms by different methods([25-30]). Among these some method([30])
is very complex but not essence and natural. But we must point out
that it is very easy to give all traveling wave solutions to
Sine-Gordon equation. In the following, we only need elementary
integral method to do this thing. Instituting traveling wave
transformation $ u=u(\xi), \xi=kx+\omega t $ into Eq.(\ref{SG}) and
integrating once yield
\begin{equation}\label{SGI}
\pm(\xi-\xi_0)=\int
\frac{\mathrm{d}u}{\sqrt{-\frac{a}{k\omega}(\cos u-c)}},
\end{equation}
where $ c $ is an arbitrary constant. Take variable transformation
\begin{equation}
u=\arccos w,
\end{equation}
then Eq.(\ref{SGI}) becomes
\begin{equation}\label{SI}
\pm(\xi-\xi_0)=\int
\frac{\mathrm{d}w}{\sqrt{\frac{a}{k\omega}(w-1)(w+1)(w-c)}}.
\end{equation}

Firstly we  assume $\frac{a}{k\omega}>0.$  There are the following
three cases to be discussed:\\

Case 1: $ c=1, $ then we have
\begin{equation}
u=arcsin\{2\tanh^2(\frac{1}{2}\sqrt{\frac{2a}{k\omega}}(\xi-{\xi}_0))-1\};
\end{equation}
\begin{equation}
u=\arcsin\{2\coth^2(\frac{1}{2}\sqrt{\frac{2a}{k\omega}}(\xi-{\xi}_0))-1\}.
\end{equation}

Case 2: $ c=-1, $ then we have
\begin{equation}
u=\arcsin\{-2\sec^2(\frac{1}{2}\sqrt{\frac{2a}{k\omega}}(\xi-{\xi}_0))+1\}.
\end{equation}

Case 3: $c\neq\pm 1, $ we reorder $ 1, -1 ,c $ and  denote them by
$ \alpha<\beta<\gamma $. In fact there are three case such as $
1>c>-1, c>1, $ and $ -1>c $.  When $ \alpha<w<\beta $, we have
\begin{equation}
u=\arcsin\{\alpha+(\beta-\alpha)\SN^2(\frac{\sqrt{\gamma-\alpha}}{2}\sqrt{\frac{2a}{k\omega}}(\xi-{\xi}_0)),m)\}.
\end{equation}\\
When $ w>\gamma $, we have
\begin{equation}
u=\arcsin\{\frac{\gamma-\beta\SN^2(\frac{\sqrt{\gamma-\alpha}}{2}\sqrt{\frac{2a}{k\omega}}(\xi-{\xi}_0)),m)}{\CN^2
(\frac{\sqrt{\gamma-\alpha}}{2}\sqrt{\frac{2a}{k\omega}}(\xi-{\xi}_0)),m)}\},
\end{equation}\\
where $ m^2=\frac{\beta-\alpha}{\gamma-\alpha} $.\\

If $\frac{a}{k\omega}<0,$ we take $ w=-v, $ then Eq.(\ref{SI})
becomes
\begin{equation}\label{S2}
\pm(\xi-\xi_0)=\int
\frac{\mathrm{d}v}{\sqrt{-\frac{a}{k\omega}(v-1)(v+1)(v+c)}}.
\end{equation}
We have similar three cases:

Case 1: $ c=1, $ then we have
\begin{equation}
u=-\arcsin\{2\tanh^2(\frac{1}{2}\sqrt{-\frac{2a}{k\omega}}(\xi-{\xi}_0))-1\};
\end{equation}
\begin{equation}
u=-\arcsin\{2\coth^2(\frac{1}{2}\sqrt{-\frac{2a}{k\omega}}(\xi-{\xi}_0))-1\}.
\end{equation}

Case 2: $ c=-1, $ then we have
\begin{equation}
u=-\arcsin\{-2\sec^2(\frac{1}{2}\sqrt{-\frac{2a}{k\omega}}(\xi-{\xi}_0))+1\}.
\end{equation}

Case 3: $c\neq\pm 1, $ we reorder $ 1, -1 ,c $ and  denote them by
$ \alpha<\beta<\gamma $. When $ \alpha<w<\beta $, we have
\begin{equation}
u=-\arcsin\{\alpha+(\beta-\alpha)\SN^2(\frac{\sqrt{\gamma-\alpha}}{2}\sqrt{-\frac{2a}{k\omega}}(\xi-{\xi}_0)),m)\}.
\end{equation}\\
When $ w>\gamma $, we have
\begin{equation}
u=-\arcsin\{\frac{\gamma-\beta\SN^2(\frac{\sqrt{\gamma-\alpha}}{2}\sqrt{-\frac{2a}{k\omega}}(\xi-{\xi}_0)),m)}{\CN^2(\frac{\sqrt{\gamma-\alpha}}{2}\sqrt{-\frac{2a}{k\omega}}(\xi-{\xi}_0)),m)}\},
\end{equation}\\
where $ m^2=\frac{\beta-\alpha}{\gamma-\alpha} $.\\

\textbf{Example 3}. Classification of all single envelope traveling
wave
solutions to Hirota equation\\

Hirota equation reads

\begin{equation}\label{HIR}
\mathrm{i}u_t+u_{xx}+2|u|^2u+\mathrm{i}\alpha
u_{xxx}+6\mathrm{i}\alpha |u|^2u_x=0,
\end{equation}
when $\alpha=0$, it becomes Schrodinger equation. Take envelope
travelling wave transformation $ u=u(\xi)\exp(\eta), \xi=x+\omega
t, \eta=px +qt$, then separating the real part and the imagine
part of Eq.(\ref{HIR}) yields
\begin{equation}\label{RE}
(\omega+2p-3\alpha p^2)u'+6\alpha u^2u'+\alpha u'''=0,
\end{equation}
\begin{equation}\label{IM}
(\alpha p^3-q-p^2)u+2(1-3\alpha p)u^2+(1-3\alpha p)u''=0.
\end{equation}
In order to make the Eq.(\ref{RE}) and the Eq.(\ref{IM})
consistence, the parameters must satisfy the following condition:
\begin{equation}\label{CON}
\frac{\omega+2p-3\alpha p^2}{\alpha
p^3-q-p^2}=\frac{\alpha}{1-3\alpha p}.
\end{equation}
Under the condition (\ref{CON}), the above two equations become
one equation
\begin{equation}
(u')^2=-u^4+\frac{\alpha p^3-q-p^2}{3\alpha p-1}u^2+c
\end{equation}
where $c$ is an arbitrary constant. The corresponding integral
form is
\begin{equation}
\pm(\xi-\xi_0)=\int\frac{\mathrm{d}u}{\sqrt{-u^4+\frac{\alpha
p^3-q-p^2}{3\alpha p-1}u^2+c}},
\end{equation}
in order to solve the above integral, we take the change of
variable $w=u^2$, then the integral becomes
\begin{equation}
\pm2(\xi-\xi_0)=\int\frac{\mathrm{d}w}{\sqrt{w(-w^2+\frac{\alpha
p^3-q-p^2}{3\alpha p-1}w+c)}},
\end{equation}
Therefore according to the classification of all solutions of the
Eq.(15), it is easy to give classification of all single envelope
traveling wave solutions of the Hirota equation. We omit concrete
expressions for briefly.

\section{Applications of the complete discrimination system for the fourth order polynomial}

We first write the complete discrimination system for the fourth
order polynomial $f(w)=w^4+pw^2+qw+r$ as follows(\cite{L3}):
\begin{equation*}
\begin{split}
D_1=4, D_2=-p, D_3=8rp-2p^3-9q^2,\\
D_4=4p^4r-p^3q^2+36prq^2-32r^2p^2 -\frac{27}{4}q^4+64r^3,
\end{split}
\end{equation*}
\begin{equation}
F_2=9p^2-32pr.
\end{equation}

Then we consider the following ODE
\begin{equation}\label{TE}
(w'(\xi))^2=\epsilon(w^4+pw^2+qw+r),
\end{equation}
where $ \epsilon=\pm1. $ Rewrite Eq.(\ref{TE}) by integral form as
follows:
\begin{equation}\label{INT}
\pm(\xi-\xi_0)=\int
\frac{\mathrm{d}w}{\sqrt{\epsilon(w^4+pw^2+qw+r)}}.
\end{equation}
Although Eq.(52) has been studied extensively, some solutions were
obtained under some special parameter's conditions. But we will
give in the following the classification of its all solutions.
According to the complete discrimination system for polynomial $
F(w)=w^4+pw^2+qw+r $, there are nine cases to be
discussed(\cite{L3}),
we list those results for convenience in the following.\\

Case 1: $ D_4=0, D_3=0, D_2<0 $. Then we have
\begin{equation}
F(w)=((w-l_1)^2+s_1^2)^2,
\end{equation}
where $ l_1, s_1 $ are real numbers, $ s_1>0 $. When $ \epsilon=1
$, we have
\begin{eqnarray}
w=s_1\tan [s_1(\xi-\xi_0)]+l_1.
\end{eqnarray}\\

Case 2: $ D_4=0, D_3=0, D_2=0 $. Then we have
\begin{equation}
F(w)=w^4.
\end{equation}
When $ \epsilon=1 $, we have
\begin{eqnarray}
w=-\frac{1}{\xi-\xi_0}.
\end{eqnarray}\\

Case 3: $ D_4=0, D_3=0, D_2>0, E_2=0 $. Then we have
\begin{equation}
F(w)=(w-\alpha)^2(w-\beta)^2,
\end{equation}
where $ \alpha, \beta$ are real numbers, $ \alpha>\beta $. If $
\epsilon=1 $, when $ w>\alpha $ or $ w<\beta, $ we have
\begin{eqnarray}
w=\frac{\beta-\alpha}{2}[\coth\frac{\alpha-\beta}{2}(\xi-\xi_0)-1]+\beta.
\end{eqnarray}
When $ \alpha>w>\beta, $ we have
\begin{eqnarray}
w=\frac{\beta-\alpha}{2}[\tanh\frac{\alpha-\beta}{2}(\xi-\xi_0)-1]+\beta.
\end{eqnarray}\\

Case 4: $ D_4=0, D_3>0, D_2>0 $. Then we have
\begin{equation}
F(w)=(w-\alpha)^2(w-\beta)(w-\gamma),
\end{equation}
where $ \alpha, \beta, \gamma $ are real numbers, and $
\beta>\gamma.$ If $ \epsilon=1$, when $ \alpha>\beta $ and $
w>\beta, $ or when $ \alpha<\gamma$ and $ w<\gamma, $ we have
\begin{eqnarray}
\pm(\xi-\xi_0)=\frac{1}{\sqrt{(\alpha-\beta)(\alpha-\gamma)}}\ln{\frac{[\sqrt{(w-\beta)(\alpha-\gamma)}-\sqrt{(\alpha-\beta)(w-\gamma)}]^2}{|w-\alpha|}}.
\end{eqnarray}
When $ \alpha>\beta $ and $ w<\gamma, $ or when $ \alpha<\gamma$
and $ w<\beta, $ we have
\begin{eqnarray}
\pm(\xi-\xi_0)=\frac{1}{\sqrt{(\alpha-\beta)(\alpha-\gamma)}}\ln{\frac{[\sqrt{(w-\beta)(\gamma-\alpha)}-\sqrt{(\beta-\alpha)(w-\gamma)}]^2}{|w-\alpha|}}.
\end{eqnarray}
When $ \beta>\alpha>\gamma, $ we have
\begin{eqnarray}
\pm(\xi-\xi_0)=\frac{1}{\sqrt{(\beta-\alpha)(\alpha-\gamma)}}\arcsin{\frac{(w-\beta)(\alpha-\gamma)+(\alpha-\beta)(w-\gamma)}{|(w-\alpha)(\beta-\gamma)|}}.
\end{eqnarray}\\

If $ \epsilon=-1, $ when $ \alpha>\beta $ and $ w>\beta, $ or when
$ \alpha<\gamma$ and $ w<\gamma, $ we have
\begin{eqnarray}
\pm(\xi-\xi_0)=\frac{1}{\sqrt{(\alpha-\beta)(\alpha-\gamma)}}\ln{\frac{[\sqrt{(w-\beta)(\alpha-\gamma)}-\sqrt{(\alpha-\beta)(w-\gamma)}]^2}{|w-\alpha|}}.
\end{eqnarray}
When $ \alpha>\beta $ and $ w<\gamma, $ or when $ \alpha<\gamma$
and $ w<\beta, $ we have
\begin{eqnarray}
\pm(\xi-\xi_0)=\frac{1}{\sqrt{(\alpha-\beta)(\alpha-\gamma)}}\ln{\frac{[\sqrt{(w-\beta)(\gamma-\alpha)}-\sqrt{(\beta-\alpha)(w-\gamma)}]^2}{|w-\alpha|}}.
\end{eqnarray}
When $ \beta>\alpha>\gamma, $ we have
\begin{eqnarray}
\pm(\xi-\xi_0)=\frac{1}{\sqrt{(\beta-\alpha)(\alpha-\gamma)}}\arcsin{\frac{(w-\beta)(\alpha-\gamma)+(\alpha-\beta)(w-\gamma)}{|(w-\alpha)(\beta-\gamma)|}}.
\end{eqnarray}\\

Case 5: $ D_4=0, D_3=0, D_2>0, E_2=0 $. Then we have
\begin{equation}
F(w)=(w-\alpha)^3(w-\beta),
\end{equation}
where $ \alpha, \beta $ are real numbers. If $ \epsilon=1, $ when
$ w>\alpha, w>\beta, $ or $ w<\alpha, w<\beta, $ we have
\begin{equation}
w=\frac{4(\alpha-\beta)}{(\alpha-\beta)^2(\xi-\xi_0)^2-4}.
\end{equation}
If $ \epsilon=-1, $ when $ w>\alpha, w<\beta; $ or $ w<\alpha,
w>\beta, $ we have
\begin{equation}
w=\frac{4(\beta-\alpha)}{4+(\alpha-\beta)^2(\xi-\xi_0)^2}.
\end{equation}\\

Case 6: $ D_4=0, D_2D_3<0 $. Then we have
\begin{equation}
F(w)=(w-\alpha)^2((w-l_1)^2+s_1^2),
\end{equation}
where $ \alpha, l_1 $ and $ s_1 $ are real numbers. If $
\epsilon=1, $ we have
\begin{eqnarray}
w=\frac{\exp{[\pm\sqrt{(\alpha-l_1)^2+s_1^2}(\xi-\xi_0)]}-\gamma+\sqrt{(\alpha-l_1)^2+s_1^2}}{(\exp{[\pm\sqrt{(\alpha-l_1)^2+s_1^2}(\xi-\xi_0)]}-\gamma)^2-1},
\end{eqnarray}
where
\begin{equation}
 \gamma=\frac{\alpha-2l_1}{\sqrt{(\alpha-l_1)^2+s_1^2}}.
\end{equation}\\

Case 7: $ D_4>0, D_3>0, D_1>0 $. Then we have
\begin{equation}
F(w)=(w-\alpha_1)(w-\alpha_2)(w-\alpha_3)(w-\alpha_4),
\end{equation}
where $ \alpha_1, \alpha_2, \alpha_3, \alpha_4 $ are real numbers,
and $\alpha_1>\alpha_2> \alpha_3>\alpha_4.$ If $\epsilon=1 $, when
$ w>\alpha_1 $ or $ w<\alpha_4, $ we have
\begin{equation}
w=\frac{\alpha_2(\alpha_1-\alpha_4)\SN^2{(\frac{\sqrt{(\alpha_1-\alpha_3)(\alpha_2-\alpha_4)}}{2}(\xi-\xi_0),m)-\alpha_1(\alpha_2-\alpha_4)}}
{(\alpha_1-\alpha_4)\SN^2{(\frac{\sqrt{(\alpha_1-\alpha_3)(\alpha_2-\alpha_4)}}{2}(\xi-\xi_0),m)-(\alpha_2-\alpha_4)}}.
\end{equation}
When $ \alpha_2>w>\alpha_3, $ we have
\begin{equation}
w=\frac{\alpha_4(\alpha_2-\alpha_3)\SN^2{(\frac{\sqrt{(\alpha_1-\alpha_3)(\alpha_2-\alpha_4)}}{2}(\xi-\xi_0),m)-\alpha_3(\alpha_2-\alpha_4)}}
{(\alpha_2-\alpha_3)\SN^2{(\frac{\sqrt{(\alpha_1-\alpha_3)(\alpha_2-\alpha_4)}}{2}(\xi-\xi_0),m)-(\alpha_2-\alpha_4)}}.
\end{equation}
If $ \epsilon=-1, $ when $ \alpha_1>w>\alpha_2, $ we have
\begin{equation}
w=\frac{\alpha_3(\alpha_1-\alpha_2)\SN^2{(\frac{\sqrt{(\alpha_1-\alpha_3)(\alpha_2-\alpha_4)}}{2}(\xi-\xi_0),m)-\alpha_2(\alpha_1-\alpha_3)}}
{(\alpha_1-\alpha_2)\SN^2{(\frac{\sqrt{(\alpha_1-\alpha_3)(\alpha_2-\alpha_4)}}{2}(\xi-\xi_0),m)-(\alpha_1-\alpha_3)}}.
\end{equation}
When $ \alpha_3>w>\alpha_4, $ we have
\begin{equation}
w=\frac{\alpha_1(\alpha_3-\alpha_4)\SN^2{(\frac{\sqrt{(\alpha_1-\alpha_3)(\alpha_2-\alpha_4)}}{2}(\xi-\xi_0),m)-\alpha_4(\alpha_3-\alpha_1)}}
{(\alpha_3-\alpha_4)\SN^2{(\frac{\sqrt{(\alpha_1-\alpha_3)(\alpha_2-\alpha_4)}}{2}(\xi-\xi_0),m)-(\alpha_3-\alpha_1)}},
\end{equation}
where $ m^2=\frac{(\alpha_1-\alpha_2)( \alpha_3-
\alpha_4)}{(\alpha_1- \alpha_3)( \alpha_2- \alpha_4)} $.\\

Case 8: $ D_4<0, D_2D_3\geq0 $. Then we have
\begin{equation}
F(w)=(w-\alpha)(w-\beta)((w-l_1)^2+s_1^2),
\end{equation}
where $ \alpha, \beta, l_1 $ and $ s_1 $ are real numbers, and $
\alpha>\beta>, s_1>0 $, we have (here positive sign for $
\epsilon=-1, $ negative for $ \epsilon=1. $)
\begin{eqnarray}
w=\frac{a\CN{(\frac{\sqrt{\mp2s_1m_1(\alpha-\beta)}}{2mm_1}(\xi-\xi_0),m)}+b}
{c\CN{(\frac{\sqrt{\mp2s_1m_1(\alpha-\beta)}}{2mm_1}(\xi-\xi_0),m)}+d}.
\end{eqnarray}
where
\begin{eqnarray*}
c=\alpha-l_1-\frac{s_1}{m_1}, d=\alpha-l_1-s_1m_1,\\
a=\frac{1}{2}[(\alpha+\beta)c-(\alpha-\beta)d],
b=\frac{1}{2}[(\alpha+\beta)d-(\alpha-\beta)c],
\end{eqnarray*}
\begin{equation}
E=\frac{s_1^2+(\alpha-l_1)(\beta-l_1)}{s_1(\alpha-\beta)}, \ \
m_1=E\pm\sqrt{E^2+1}, m^2=\frac{1}{1+m_1^2},
\end{equation}
we choose $ m_1 $ such that $ \epsilon m_1<0. $\\

Case 9: $ D_4>0, D_2D_3\leq0, $ then we have
\begin{equation}
F(w)=((w-l_1)^2+s_1^2)(w-l_2)^2+s_2^2),
\end{equation}
where $ l_1, l_2, s_1 $ and $ s_2 $ are real numbers, and $
s_1>s_2>0 $. If $ \epsilon=1, $  we have
\begin{equation}
w=\frac{a\SN{(\eta(\xi-\xi_0),m)}+b\CN{(\eta(\xi-\xi_0),m)}}{c\SN{(\eta(\xi-\xi_0),m)}+d\CN{(\eta(\xi-\xi_0),m)}},
\end{equation}
where
\begin{eqnarray*}
c=-s_1-\frac{s_2}{m_1}, \ \ d=l_1-l_2, \ \  a=l_1c+s_1d, \ \ b=l_1d-s_1c,\\
E=\frac{s_1^2+s_2^2+(l_1-l_2)^2}{2s_1s_2}, \ \ m_1=E+\sqrt{E^2-1},
\end{eqnarray*}
\begin{equation}
m^2=1-\frac{1}{m_1^2}, \ \
\eta=s_2\sqrt{\frac{m_1^2c^2+d^2}{c^2+d^2}}.
\end{equation}\\

\textbf{Remark 1}: For case 7 and case 8, we can use another
transformation to obtain more simple representations. For example,
for case 7, we consider only $ \epsilon=1,$ as an example, we take
a transformation as follows:
\begin{equation}\label{T6}
w_1=\frac{A^{\frac{1}{3}}}{w-\alpha_1},
\end{equation}
where $
A=(\alpha_1-\alpha_2)(\alpha_1-\alpha_3)(\alpha_1-\alpha_4),$ then
corresponding integral becomes
\begin{equation}
 \pm a^{\frac{1}{3}}(\xi-\xi_0)=\int \frac{\mathrm{d}w_1}{\sqrt{(w_1-\beta_1)(w_1-\beta_2)(w_1-\beta_3))}}.
\end{equation}
where $\beta_1=\frac{A^{\frac{1}{3}}}{\alpha_4-\alpha_1}, \ \
\beta_2=\frac{A^{\frac{1}{3}}}{\alpha_3-\alpha_1}, \ \
\beta_3=\frac{A^{\frac{1}{3}}}{\alpha_2-\alpha_1}$.  when $
\beta_1>w_1>\beta_2, $ we have
\begin{equation}
w_1=\beta_3+(\beta_2-\beta_3)\SN^2{(\frac{\sqrt{\beta_1-\beta_3}}{2}(\xi-\xi_0),m)}.
\end{equation}
When $ w_1>\beta_1, $ we have
\begin{equation}
w_1=\frac{\beta_1-\beta_2\SN{(\frac{\sqrt{\beta_1-\beta_3}}{2}(\xi-\xi_0),m)}}
{\CN{(\frac{\sqrt{\beta_1-\beta_3}}{2}(\xi-\xi_0),m)}}.
\end{equation}
where $ m=\frac{\beta_2-\beta_3}{\beta_1-\beta_3}.$\\

For the case of $ \epsilon=-1 $ and the case 8, we can deal with them similarly.\\

\textbf{Remark 2}: All above integrals can be fond in Ref.\cite{ZLZ}.\\

\textbf{Example 1}. Classification of all single travelling wave solutions to mKdV equation.\\

mKdV equation reads
\begin{equation}\label{MK}
u_t+au^2u_x+bu_{xxx}=0.
\end{equation}
Its exact traveling wave solutions have been studied
extensively([20-23]). But its all traveling wave solutions haven't
been reported. Using direct integral method and complete
discrimination system for the fourth order polynomial, we give in
the following all traveling wave solutions to mKdV equation.

Take traveling wave transformation as follows:
\begin{equation}\label{WT}
u=u(\xi), \xi=kx+\omega t.
\end{equation}
Instituting Eq.(\ref{WT}) into Eq.(\ref{MK}) and integrating twice
yield
\begin{equation}\label{In}
(u')^2=A(u^4+pu^2+qu+r),
\end{equation}
where
\begin{equation}
A=-\frac{a}{6bk^2}, \ \ p=\frac{ak}{6\omega},
\end{equation}
and $ q, r $ are two arbitrary constants. Write Eq.(\ref{In}) by
integral form
\begin{equation}\label{INT}
\pm\sqrt{\epsilon A}(\xi-\xi_0)=\int
\frac{\mathrm{d}u}{\sqrt{\epsilon(u^4+pu^2+qu+r)}},
\end{equation}
where $ \epsilon=\pm1. $ If $ A>0, $ we take $ \epsilon=1.$ If $
A<0,$ we take $ \epsilon=-1.$ According to the former results, we
can give the classification of all single traveling wave
solutions to MKdV equation, we omit them for briefly.\\

\textbf{Example 2}. Classification of all single traveling solutions
to combined KdV-MKdV equation(\cite{KM1},\cite{FAN3, \cite{CKM2}})
which reads
\begin{equation}\label{MKDV}
u_t+auu_x+bu^2u_x+u_{xxx}=0,
\end{equation}
where $ a $ and $ b $ are physical parameters. Taking traveling wave
transformation $ u=u(\xi_1), \xi_1=kx-\omega t, $ and instituting it
into Eq.(\ref{MKDV}) and integrating once yields:
\begin{equation}\label{MI}
(u')^2=a_4u^4+a_3u^3+a_2u^2+a_1u+a_0,
\end{equation}
where
\begin{equation}
a_4=-\frac{a}{6k^2}, \ \ a_3=-\frac{b}{3k^2}, \ \
a_2=\frac{\omega}{k^3},
\end{equation}
and $ a_1, a_0 $ are arbitrary constants. If $ a<0, $ we take
transformations as follows:
\begin{equation}
w=a_4^{\frac{1}{4}}(u+\frac{a_3}{4a_4}), \ \
\xi=a_4^{\frac{1}{4}}\xi_1.
\end{equation}
Then Eq.(\ref{MI}) becomes
\begin{equation}
(w')^2=w^4+pw^2+qu+r,
\end{equation}
where
\begin{equation}
p=\frac{a_2}{\sqrt{a_4}},\ \
q=a_4^{-\frac{1}{4}}(\frac{a_3^3}{8a_4^2}-\frac{a_2a_3}{2a_4}+a_1),
\ \
r=a_0-\frac{a_1a_3}{4a_4}+\frac{a_2a_3^2}{16a_4^2}-\frac{3a_3^4}{256a_4^3}.
\end{equation}\\

If $ a>0, $ we take transformations as follows:
\begin{equation}
w=(-a_4)^{\frac{1}{4}}(u+\frac{a_3}{4a_4}), \ \
\xi=(-a_4)^{\frac{1}{4}}\xi_1
\end{equation}
Then Eq.(\ref{MI}) becomes
\begin{equation}
(w')^2=-(w^4+pw^2+qu+r),
\end{equation}
where
\begin{equation}
p=-\frac{a_2}{\sqrt{-a_4}},\ \
q=(-a_4)^{-\frac{1}{4}}(-\frac{a_3^3}{8a_4^2}+\frac{a_2a_3}{2a_4}-a_1),
\ \
r=-a_0-\frac{a_1a_3}{4a_4}-\frac{a_2a_3^2}{16a_4^2}+\frac{3a_3^4}{256a_4^3}.
\end{equation}\\

According to the classification of solutions to ODE $
(w')^2=\pm(w^4+pw^2+qw+r)$, we can give all traveling wave
solutions to combined KdV-MKdV equation. We omit them for simplicity\\

\textbf{Example 3}. Classification of all single traveling wave
solutions to Konopelchenko-Dubrovsky equation (\cite{GH, KDP}) which
reads
\begin{eqnarray}\label{KD}
u_t=u_{xxx}-6\beta uu_x+\frac{3}{2}\alpha^2u^2u_x-3w_y+3\alpha uu_x=0,\\
w_x=u_y.
\end{eqnarray}
Takeing travelling wave transformation $ u=u(\xi),
\xi=k_1x+k_2y+\omega t, $ instituting it into Eq.(\ref{KD}) and
integrating twice yield the same form equation with Eq.(\ref{IN}):
\begin{equation}
(u')^2=a_4u^4+a_3u^3+a_2u^2+a_1u+a_0,
\end{equation}
where
\begin{equation}
a_4=\frac{\alpha^2}{4k_1^2}, \ \ a_3=\frac{\alpha-2\beta}{2k_1^2},
\ \ a_2=\frac{2(k_1\omega-3k_2^2)}{k_1^4},
\end{equation}
and $ a_1, a_0 $ are arbitrary constants. Notice $a_4>0$. According
to the example 1, it is easy to write all traveling wave solutions
to KD equation, we omit them for briefly.
All nine cases under the condition $a_4>0$ in section 4 are possible.   \\

\textbf{Example 4}: Classification of all single traveling solutions
to double Sine-Gordon equation (\cite{LAM, BCG}) which reads
\begin{equation}\label{DSG}
u_{xx}-u_{tt}=\sin{au}+\lambda\sin{\frac{a}{2}u},
\end{equation}\\
where $ a $ and $ \lambda $ are nonzero constants. Its some exact
traveling wave solutions have been obtained. We give in the
following its all traveling wave solutions.\\

Take traveling wave transformation
\begin{equation}\label{Wa}
u=u(\xi_1), \ \ \xi_1=kx-{\omega}t, \ \ k\neq\pm\omega.
\end{equation}\\
Instituting Eq.(\ref{Wa}) into Eq.(\ref{DSG}) and integrating once
yields
\begin{equation}\label{IN}
\pm(\xi_1-\xi_{10})=\int
\frac{\mathrm{d}u}{\sqrt{c_2-c_1\cos{au}-c_0\cos{\frac{a}{2}u}}}.
\end{equation}\\
where
\begin{equation}
c_1=\frac{2}{a(k_2-\omega^2)},
c_0=\frac{4\lambda}{a(k_2-\omega^2)}.
\end{equation}\\
In order to solve above integral (\ref{IN}), we take
transformation as follows:
\begin{equation}\label{TH}
u=\frac{2}{a}\arccos{v}.
\end{equation}\\
Instituting Eq.(\ref{TH}) into integral (\ref{IN}), we have
\begin{equation}\label{INN}
\pm(\xi_1-\xi_{10})=\int
\frac{\frac{2}{a}\mathrm{d}v}{\sqrt{(v-1)(v+1)(2c_1v^2+c_0v+c_1-c_2)}}.
\end{equation}\\
If $ c_1>0, $ let
\begin{equation}\label{W}
w=\sqrt{c_1}v+\frac{c_0}{4c_1}, \ \
\xi=\frac{a}{\sqrt{2c_1}}\xi_1.
\end{equation}
Thus Eq.(\ref{INN}) becomes respectively
\begin{equation}
\pm(\xi-\xi_0)=\int
\frac{\mathrm{d}v}{\sqrt{(w-\alpha)(w-\beta)(w^2+d_0)}}.
\end{equation}
where
\begin{eqnarray}
\alpha=\frac{c_0}{4c_1}+\sqrt{c_1}, \ \
\beta=\frac{c_0}{4c_1}-\sqrt{c_1}, \ \
d_0=\frac{c_1-c_2}{2}-\frac{c_0^2}{14c_1^2}.
\end{eqnarray}\\
If $ c_1<0, $ let
\begin{equation}\label{W}
w=\sqrt{-c_1}v-\frac{c_0}{4c_1}, \ \
\xi=\frac{a}{\sqrt{2c_1}}\xi_1.
\end{equation}
Thus Eq.(\ref{INN}) becomes respectively
\begin{equation}
\pm(\xi-\xi_0)=\int
\frac{\mathrm{d}v}{\sqrt{-(w-\alpha)(w-\beta)(w^2+d_0)}}.
\end{equation}
where
\begin{eqnarray}
\alpha=-\frac{c_0}{4c_1}+\sqrt{-c_1}, \ \
\beta=-\frac{c_0}{4c_1}-\sqrt{-c_1}, \ \
d_0=\frac{c_1-c_2}{2}-\frac{c_0^2}{14c_1^2}.
\end{eqnarray}\\

Denote
\begin{equation}\label{FW1}
F(w)=(w-\alpha)(w-\beta)(w^2+d_0),
\end{equation}\\
and rewrite it as follows:
\begin{equation}
F(w)=w^4+pw^2+qw+r,
\end{equation}\\
where
\begin{eqnarray}
p=d_0-(\alpha+\beta), \ \ r=-d_0(\alpha+\beta), \ \
s=d_0\alpha\beta.
\end{eqnarray}\\
But because of special form of $ F(w) $ (see Eq.(\ref{FW1})), we
know that the case 1, case 2, case 6 and case 9 in section 3 are
impossible to occur in the case of double Sine-Gordon equation.
Other cases are possible. Thus we can give all traveling solutions
to double
Sine-Gordon equation according to the cases 3-5 and cases 7-8 in section 3. We omit them for briefly.\\

\textbf{Example 5}. Classification of all single traveling wave
solutions to Pochharmer-Chree Equation (\cite{SE}) which reads
\begin{equation}\label{PC}
u_{tt}-\alpha u_{xx}+\beta u+\gamma u^3+\delta u^5.
\end{equation}
Takeing traveling wave transformation $ u=u(\xi), \xi=kx+\omega t, $
instituting it into Eq.(\ref{PC}) and integrating once yields
\begin{equation}\label{PCT}
(u')^2=a_3u^6+a_2u^4+a_1u^2+a_0,
\end{equation}
where
\begin{equation}
a_3=-\frac{\delta}{3(\omega^2-\alpha k^2)}, \ \
a_2=-\frac{\gamma}{2(\omega^2-\alpha k^2)}, \ \
a_1=-\frac{\beta}{\omega^2-\alpha k^2},
\end{equation}
and $ a_0 $ is an arbitrary constant. Let $ v=u^2, $ then
Eq.(\ref{PCT}) becomes
\begin{equation}\label{VE}
(v')^2=2v(a_3v^3+a_2v^2+a_1v+a_0).
\end{equation}
Similar to the example 3, we can give all traveling wave solutions
to Pochhamer-Chree  equation (\ref{PC}). Because of special form of
the right polynomial of Eq.(\ref{VE}), the case 1 and case 9 in
section 3 are
impossible to occur, all other cases 2-8 are possible. We omit them for simplicity.\\

\textbf{Example 6}: Classification of all single  traveling wave
solutions to classical Bousinesq system
\begin{equation}
\eta_t+[(1+\eta)u]_x=-\frac{1}{4}u_{xxx},
\end{equation}
\begin{equation}
u_t+uu_x\eta_x=0.
\end{equation}
Take traveling wave transformation $ \eta=\eta(\xi), u=u(\xi),
\xi=kx+\omega t$, instituting it into above system and integrating
once yield
\begin{equation}\label{e1}
c_0+\omega\eta+k(1+\eta)u=-\frac{k^3}{4}u'',
\end{equation}
\begin{equation}\label{e2}
\eta=c_1-\frac{\omega}{k} u-\frac{1}{2}u^2.
\end{equation}
Instituting Eq.(\ref{e2}) into Eq.(\ref{e1}) and integrating
once,we have
\begin{equation}
(u')^2=a_4u^4+a_3u^3+a_2u^2+a_1u+a_0,
\end{equation}
where
\begin{equation}
 a_4=\frac{1}{k^2}, \ \ a_3=\frac{8\omega}{3k^3}, \ \ a_2=\frac{4\omega^2-4k^2(1+c_1)}{k^4}, \ \
 a_1=c_0+c_1\omega,
 \end{equation}
 and $a_0 $ is an arbitrary constant. Notice  $a_4>0$ and according to the example 1, we can easily to obtain
 all traveling wave solutions to classical Boussinesq system. For briefly we omit them.

\section{Applications of the complete discrimination system for the fifth order polynomial }

We consider the following ODE
\begin{equation}\label{St}
(w'(\xi))^2=F(w)=w^5+pw^3+qw^2+rw+s.
\end{equation}
We write its complete discrimination system as follows (see
\cite{YH,L2}):
\begin{equation*}
\begin{split}
D_2=-p, D_3=40rp-12p^3-45q^2,\\
D_4=12p^4r-4p^3q^2+117prq^2-88r^2p^2\\
-40qsp^2-27q^4-300qrs+160r^3,\\
D_5=-1600qsr^3-3750pqs^3+2000ps^2r^2-4p^3q^2r^2+16p^3q^3s\\-900rs^2p^3
+825p^2q^2s^2+144pq^2r^3+2250rq^2s^2+16p^4r^3\\+108p^5s^2-128r^4p^2
-27r^2q^4+108sq^5+256r^5+3125s^4\\-72rsqp^4+560sqr^2p^2-630prsq^3,\\
\end{split}
\end{equation*}
\begin{equation*}
\begin{split}
E_2=160r^2p^3+900q^2r^2-48rp^5+60rP^2q^2+1500pqrs+16q^2p^4\\-1100qsp^3
+625s^2p^2-3375sq^3,
\end{split}
\end{equation*}
\begin{equation}\label{DSC4}
 F_2=3q^2-8rp.
\end{equation}

According to the complete discrimination system (\ref{DSC4}) for
polynomial $
f(w) $, there are twelve cases to be discussed.\\

Case 1: $ D_5=0, D_4=0, D_3>0, E_2\neq0 $. Then we have
\begin{equation}
F(w)=(w-\alpha)^2(w-\beta)^2(w-\gamma),
\end{equation}
where $ \alpha, \beta, \gamma $ are real numbers, $
\alpha\neq\beta\neq\gamma $. When $ w>\gamma $, we have
\begin{eqnarray}
\pm\frac{\alpha-\beta}{2}(\xi-\xi_0)=\sqrt{\gamma-\alpha}\arctan{\frac{\sqrt{w-\gamma}}{\sqrt{\gamma-\alpha}}}
-\sqrt{\gamma-\beta}\arctan{\frac{\sqrt{w-\gamma}}{\sqrt{\gamma-\beta}}},\cr\gamma>\alpha,
\gamma>\beta;
\end{eqnarray}
\begin{eqnarray}
\pm\frac{\alpha-\beta}{2}(\xi-\xi_0)=\sqrt{\gamma-\alpha}\arctan{\frac{\sqrt{w-\gamma}}{\sqrt{\gamma-\alpha}}}
-\frac{1}{\sqrt{\beta-\gamma}}\cr\times\ln{|\frac{\sqrt{w-\gamma}-\sqrt{\beta-\gamma}}{\sqrt{w-\gamma}+\sqrt{\beta-\gamma}}|},\gamma>\alpha,
\gamma<\beta;
\end{eqnarray}
\begin{eqnarray}
\pm\frac{\alpha-\beta}{2}(\xi-\xi_0)=-\sqrt{\gamma-\beta}\arctan{\frac{\sqrt{w-\gamma}}{\sqrt{\gamma-\beta}}}
+\frac{1}{2\sqrt{\alpha-\gamma}}\cr\times\ln{|\frac{\sqrt{w-\gamma}-\sqrt{\alpha-\gamma}}{\sqrt{w-\gamma}+\sqrt{\alpha-\gamma}}|},
\gamma<\alpha, \gamma>\beta;
\end{eqnarray}
\begin{eqnarray}
\pm\frac{\alpha-\beta}{2}(\xi-\xi_0)=\frac{1}{2\sqrt{\alpha-\gamma}}\ln{|\frac{\sqrt{w-\gamma}-\sqrt{\alpha-\gamma}}{\sqrt{w-\gamma}+\sqrt{\alpha-\gamma}}|}
-\frac{1}{2\sqrt{\beta-\gamma}}\cr\times\ln{|\frac{\sqrt{w-\gamma}-\sqrt{\beta-\gamma}}{\sqrt{w-\gamma}+\sqrt{\beta-\gamma}}|},\gamma<\alpha,
\gamma<\beta.
\end{eqnarray}\\

Case 2: $ D_5=0, D_4=0, D_3=0, D_2\neq0, F_2\neq0 $. Then we have
\begin{equation}
F(w)=(w-\alpha)^3(w-\beta)^2,
\end{equation}
where $ \alpha, \beta,$ are real numbers, $ \alpha\neq\beta$. When
$ w>\alpha $, we have
\begin{eqnarray}
\pm\frac{\alpha-\beta}{2}(\xi-\xi_0)=-\frac{1}{\sqrt{w-\alpha}}
-\sqrt{\alpha-\beta}\arctan{\frac{\sqrt{w-\alpha}}{\sqrt{\alpha-\beta}}},
\alpha>\beta;
\end{eqnarray}
\begin{eqnarray}
\pm\frac{\alpha-\beta}{2}(\xi-\xi_0)=-\frac{1}{\sqrt{w-\alpha}}
-\frac{1}{2\sqrt{\beta-\alpha}}\ln{|\frac{\sqrt{w-\alpha}-\sqrt{\beta-\alpha}}{\sqrt{w-\alpha}+\sqrt{\beta-\alpha}}|},
\alpha<\beta.
\end{eqnarray}\\

Case 3: $ D_5=0, D_4=0, D_3=0, D_2\neq0, F_2=0 $. Then we have
\begin{equation}
F(w)=(w-\alpha)^4(w-\beta),
\end{equation}
where $ \alpha, \beta,$ are real numbers, $ \alpha\neq\beta$. When
$ w>\alpha $, we have
\begin{eqnarray}
\pm\frac{\alpha-\beta}{2}(\xi-\xi_0)=-\frac{\sqrt{w-\beta}}{2(w-\alpha)}
-\frac{1}{2\sqrt{\alpha-\beta}}\arctan{\frac{\sqrt{w-\beta}}{\sqrt{\beta-\alpha}}},
\alpha<\beta;
\end{eqnarray}
\begin{eqnarray}
\pm\frac{\alpha-\beta}{2}(\xi-\xi_0)=-\frac{\sqrt{w-\alpha}}{2(w-\alpha)}
-\frac{1}{4\sqrt{\alpha-\beta}}\ln{|\frac{\sqrt{w-\beta}-\sqrt{\alpha-\beta}}{\sqrt{w-\beta}+\sqrt{\alpha-\beta}}|},
\alpha>\beta.
\end{eqnarray}\\

Case 4: $ D_5=0, D_4=0, D_3=0, D_2=0 $. Then we have
\begin{equation}
F(w)=(w-\alpha)^5,
\end{equation}
where $ \alpha $ is real number. When $ w>\alpha $, we have
\begin{equation}
\pm(\xi-\xi_0)=-\frac{2}{3}(w-\alpha)^{-\frac{2}{3}}.
\end{equation}\\

Case 5: $ D_5=0, D_4=0, D_3<0, E_2\neq0 $. Then we have
\begin{equation}
F(w)=(w-\alpha)(w^2+rw+s)^2,
\end{equation}
where $ \alpha $ is real number, $ r^2-4s<0 $. When $ w>\alpha $,
we have
\begin{eqnarray}
\pm(\xi-\xi_0)=-\frac{2}{\rho\sqrt{4s-r^2}}(\cos\varphi\arctan{\frac{2\rho\sin\varphi\sqrt{w-\alpha}}{w-\alpha-\rho^2}}+\frac{\sin\varphi}{2}\cr
\times\ln|\frac{w-\alpha-\rho^2-2\rho\cos\varphi\sqrt{w-\alpha}}{w-\alpha-\rho^2+2\rho\cos\varphi\sqrt{w-\alpha}}|),
\end{eqnarray}
where
\begin{equation}
\rho=(\alpha^2+r\alpha+s)^\frac{1}{4},
\varphi=\frac{1}{2}\arctan{\frac{\sqrt{4s-r^2}}{-2\alpha-r}}.
\end{equation}\\

Case 6: $ D_5=0, D_4>0 $. Then we have
\begin{equation}
F(w)=(w-\alpha)^2(w-\alpha_1)(w-\alpha_2)(w-\alpha_3),
\end{equation}
where $ \alpha, \alpha_1, \alpha_2, \alpha_3 $ are real numbers, $
\alpha_1>\alpha_2>\alpha_3. $  We have
\begin{equation}
\pm(\xi-\xi_0)=-\frac{2}{(\alpha-\alpha_2)\sqrt{\alpha_2-\alpha_3}}\{\FF(\varphi,k)
-\frac{\alpha_1-\alpha_2}{\alpha_1-\alpha}\Pi(\varphi,\frac{\alpha_1-\alpha_2}{\alpha_1-\alpha},k)\},
\end{equation}
where $ \alpha\neq\alpha_1, \alpha\neq\alpha_2, \alpha\neq\alpha_3
$ and
\begin{equation}
\FF(\varphi,k)=\int_{0}^\varphi
\frac{\mathrm{d}\varphi}{\sqrt{1-k^2\sin^2\varphi}},
\end{equation}
\begin{equation}
\Pi(\varphi,n,k)=\int_{0}^\varphi
\frac{\mathrm{d}\varphi}{(1+n\sin^2\varphi)\sqrt{1-k^2\sin^2\varphi}}.
\end{equation}\\

Case 7: $ D_5=0, D_4=0, D_3<0, E_2=0 $. Then we have
\begin{equation}
F(w)=(w-\alpha)^3((w-l_1)^2+s_1^2),
\end{equation}
where $ \alpha, l_1 $ and $ s_1 $ are real numbers. When $
w>\alpha $, if $ \alpha\neq l_1+s_1, $ we have
\begin{eqnarray}\label{C7}
\pm(\xi-\xi_0)=\frac{\tan\theta+\cot\theta}{2(s_1\tan\theta-l-\alpha)\sqrt\frac{s_1}{\sin^3{2\theta}}}\FF(\varphi,k)-
\frac{s_1\tan\theta+s_1\cot\theta}{s_1\cot\theta+l_!+\alpha}\cr
\times\{(\frac{\tan\theta+l_1+\alpha}{(s_1\cot\theta+l_1-\alpha)\sin\varphi}\sqrt{1-k^2\sin^2\varphi}
+\FF(\varphi,k)-\EE(\varphi,k)\};
\end{eqnarray}
if $ \alpha=l_1+s_1, $ we have
\begin{equation}
\pm(\xi-\xi_0)=\sqrt\frac{\sin^3{2\theta}}{4s^3}(\frac{1}{k}\arcsin(k\sin\varphi)-\FF{(\varphi,k)}),
\end{equation}
where
\begin{equation}
\tan2\theta=\frac{s_1}{\alpha-l_1}, \ \ k=\sin\theta, \ \
0<\theta<\frac{\pi}{2}, \ \ \EE{(\varphi,k)}=\int_0^\varphi
\sqrt{1-k^2\sin^2\psi}\mathrm{d}\psi.
\end{equation}\\

Case 8: $ D_5=0, D_4<0 $. Then we have
\begin{equation}
F(w)=(w-\alpha)^2(w-\beta)((w-l_1)^2+s_1^2),
\end{equation}
where $ \alpha, l_1 $ and $ s_1 $ are real numbers. When $ w>\beta
$, if $ \alpha\neq l_1-s_1\tan \theta, \alpha\neq l_1+s_1\cot
\theta, $ we have
\begin{eqnarray}
\pm(\xi-\xi_0)=\frac{\tan\theta+\cot\theta}{2(s_1\tan\theta-l-\alpha)\sqrt\frac{s_1}{\sin^3{2\theta}}}\FF(\varphi,k)-
\frac{s_1\tan\theta+s_1\cot\theta}{s_1\cot\theta+l_!+\alpha}\cr
\times\{(\frac{\tan\theta+l_1+\alpha}{(s_1\cot\theta+l_1-\alpha)\sin\varphi}\sqrt{1-k^2\sin^2\varphi}
+\FF(\varphi,k)-\EE(\varphi,k)\};
\end{eqnarray}
if $ \alpha= l_1-s_1\tan \theta $ we have
\begin{equation}
\pm(\xi-\xi_0)=\sqrt\frac{\sin^3{2\theta}}{4s^3}(\frac{1}{k}\arcsin(k\sin\varphi)-\FF{(\varphi,k)});
\end{equation}
if $ \alpha= l_1+s_1\cot \theta, $ we have
\begin{equation}
\pm(\xi-\xi_0)=\sqrt\frac{\sin^3{2\theta}}{4s^3}(\FF{(\varphi,k)}-\frac{1}{\sqrt{1-k^2}}\ln\frac{\sqrt{1-k^2\sin^2\varphi}+\sqrt{1-k^2}\sin
\varphi}{\cos \varphi});
\end{equation}
where
\begin{equation}
\tan2\theta=\frac{s_1}{\beta-l_1}, \ \ k=\sin\theta, \ \
0<\theta<\frac{\pi}{2}.
\end{equation}\\

Case 9: $ D_5=0, D_4=0, D_3>0, E_2=0 $. Then we have
\begin{equation}
F(w)=(w-\alpha)^3(w-\beta)(w-\gamma),
\end{equation}
where $ \alpha, \beta $ and $ \gamma $ are real numbers. When $
w>\alpha>\beta>\gamma $, we have
\begin{eqnarray}
\pm\frac{\alpha-\beta}{2}(\xi-\xi_0)=\frac{1}{\sqrt{\alpha-\gamma}}\EE(\arcsin\sqrt{\frac{\alpha-\gamma}{w-\gamma}},\sqrt{\frac{\beta-\gamma}{\alpha-\gamma}})
\cr-\sqrt{\frac{w-\beta}{(w-\gamma)(w-\alpha)}}.
\end{eqnarray}
In other cases such as $ w>\beta>\alpha>\gamma $ and so on, we can
give the corresponding solutions similarly. We omit them for
simplicity.\\

Case 10; $ D_5>0, D_4>0, D_3>0, D_2>0, $ we have
\begin{equation}
F(w)=(w-\alpha_1)(w-\alpha_2)(w-\alpha_3)(w-\alpha_4)(w-\alpha_5).
\end{equation}
Then the corresponding solutions can be expressed by
hyper-elliptic functions or hyper-elliptic integral such as
\begin{equation}
\pm(\xi-\xi_0)=\int
\frac{\mathrm{d}w}{\sqrt{(w-\alpha_1)(w-\alpha_2)(w-\alpha_3)(w-\alpha_4)(w-\alpha_5)}}.
\end{equation}\\

Case 11: $ D_5<0, $ we have
\begin{equation}
F(w)=(w-\alpha_1)(w-\alpha_2)(w-\alpha_3)((w-l_1)^2+s_1^2).
\end{equation}
Then the corresponding solutions can be expressed by
hyper-elliptic functions or hyper-elliptic integral such as
\begin{equation}
\pm(\xi-\xi_0)=\int
\frac{\mathrm{d}w}{\sqrt{(w-\alpha_1)(w-\alpha_2)(w-\alpha_3)((w-l_1)^2+s_1^2)}}.
\end{equation}\\

Case 12: $ D_5>0\bigwedge(D_4\leq0\bigvee D_3\leq0\bigvee
D_2\leq0), $ where $ \bigwedge $ means "and", $\bigvee$ means
"or". Then we have
\begin{equation}
F(w)=(w-\alpha)((w-l_1)^2+s_1^2))((w-l_2)^2+s_2^2).
\end{equation}
Then the corresponding solutions can be expressed by
hyper-elliptic functions or hyper-elliptic integral such as
\begin{equation}
\pm(\xi-\xi_0)=\int
\frac{\mathrm{d}w}{\sqrt{(w-\alpha)((w-l_1)^2+s_1^2))((w-l_2)^2+s_2^2)}}.
\end{equation}\\

Example 1: All traveling solutions to triple Sine-Gordon
equation([3],[42-45])
\begin{equation}\label{TR}
u_{xx}-u_{tt}=c\sin{u}+\frac{1}{3}a\sin{\frac{1}{3}u}+\frac{2}{3}b\sin{\frac{2}{3}u},
\end{equation}\\
where $ a, b $ and $ c $ are nonzero constants. Its some exact
traveling wave solutions have been obtained. In Ref.[3], we used the
similar idea to deal with it, but we didn't recognize that its all
single traveling wave solutions can be obtained. We give in
the following its all traveling wave solutions.\\

Take traveling wave transformation
\begin{equation}\label{Wa1}
\xi_1=kx-{\omega}t, \ \ k\neq\pm\omega.
\end{equation}\\
Instituting Eq.(\ref{Wa1}) into Eq.(\ref{TR}) and integrating once
yields
\begin{equation}\label{IN1}
\pm(\xi_1-\xi_{10})=\int
\frac{\mathrm{d}u}{\sqrt{c_2-2c_1\cos{u}-2a_1\cos{\frac{1}{3}u}-2b_1\cos{\frac{2}{3}u}}},
\end{equation}\\
where
\begin{equation}
a_1=\frac{a}{k_2-\omega^2},
b_1=\frac{b}{k_2-\omega^2},c_1=\frac{c}{k_2-\omega^2}.
\end{equation}\\
In order to solve above integral (\ref{IN1}), we take
transformation as follows:
\begin{equation}\label{TH1}
u=3\arccos{v}.
\end{equation}\\
Instituting Eq.(\ref{TH1}) into integral (\ref{IN1}), we have
\begin{equation}\label{IN2}
\pm(\xi_1-\xi_{10})=\int
\frac{3\mathrm{d}v}{\sqrt{(v-1)(v+1)(8c_1v^3+4b_1v^2+(2a_1-6c_1)v-2b_1-c_2)}}.
\end{equation}\\
Let
\begin{equation}\label{W}
v=(8c_1)^{-\frac{1}{5}}w-\frac{b_1}{6c_1}, \ \
\xi=\frac{1}{3}(8c_1)^{\frac{1}{5}}\xi_1.
\end{equation}\\
Thus Eq.(\ref{IN2}) becomes
\begin{equation}\label{In1}
\pm(\xi-\xi_0)=\int
\frac{\mathrm{d}v}{\sqrt{(w-\alpha)(w-\beta)(w^3+d_1w+d_0)}},
\end{equation}\\
where
\begin{eqnarray}
\alpha=(8c_1)^{\frac{1}{5}}(\frac{b_1}{6c_1}+1),
\beta=(8c_1)^{\frac{1}{5}}(\frac{b_1}{6c_1}-1);\\
d_1=(8c_1)^{\frac{2}{5}}(\frac{a_1}{4c_1}-\frac{b_1^2}{12c_1^2}-\frac{3}{4}),\
\
d_0=8c_1)^{\frac{3}{5}}(\frac{b_1^3}{108c_1^3}-\frac{b_1a_1}{24c_1^2}-\frac{(b_1+c_2)}{8c_1}).
\end{eqnarray}\\
Denote
\begin{equation}\label{FW5}
F(w)=(w-\alpha)(w-\beta)(w^3+d_1w+d_0).
\end{equation}\\
Rewrite it as follows:
\begin{equation}
F(w)=w^5+pw^3+qw^2+rw+s,
\end{equation}\\
where
\begin{eqnarray}
p=\alpha\beta-d_1(\alpha+\beta), q=d_0-d_1(\alpha+\beta),
r=\alpha\beta{d_1}-d_0(\alpha+\beta), s=d_0\alpha\beta.
\end{eqnarray}\\
But because of special form of $ F(w) $ (see Eq.(\ref{FW5})), we
know that the case 4, case 5, case 7 and case 12 in section 4 are
impossible to occur in the case of triple Sine-Gordon equation.
Other cases are possible. Thus we can give all traveling solutions
to triple Sine-Gordon equation according to the cases 1-3, case 6
and cases
8-11 in section 4. We omit them for briefly.\\

Example 2. All traveling wave solutions to $ D+1 $ dimensional
Klein-Gordon equation with fourth order nonlinear term
\begin{equation}\label{KG}
\sum_{i=1}^D u_{x_ix_i}-u_{tt}=\sum_{j=0}^4 b_ju^j.
\end{equation}\\
Take traveling wave transformation $ u=u(\xi_1), \ \
\xi_1=\sum_{i=1}^D k_ix_i+\omega{t}. $  Then Eq.(\ref{KG}) becomes
\begin{equation}\label{EQ}
(u')^2=a_0+\sum_{i=1}^5 a_iu^i,
\end{equation}\\
where $ a_0 $ is integral constant and
\begin{equation}
a_i=\frac{2b_i}{(i+1)(\sum_{i=1}^{D+1} k_i^2-\omega^2)}.
\end{equation}\\

In order to reduce Eq.(\ref{EQ}) to the form (\ref{St}), we take
transformations as follows:
\begin{equation}
w=a_5^{\frac{1}{5}}u+\frac{4}{5}a_4a_5^{-\frac{4}{5}}, \ \
\xi=a_5^{\frac{1}{5}}\xi_1.
\end{equation}\\
Then Eq.(\ref{EQ}) becomes
\begin{equation}
\pm(\xi-\xi_0)=\int \frac{\mathrm{d}w}{\sqrt{w^5+pw^3+qw^2+rw+s}},
\end{equation}\\
where
\begin{eqnarray*}
p=\frac{16}{5}a_4^2a_5^{-\frac{8}{5}}+a_3a_5^{-\frac{3}{5}},q=-\frac{32}{25}a_4^3a_5^{-\frac{12}{5}}+(a_2-3a_3)a_5^{-\frac{2}{5}},
\end{eqnarray*}
\begin{eqnarray}
r=\frac{256}{125}a_4^4(a_5^{-4}-a_5^{-\frac{16}{5}})+\frac{48}{25}a_3a_4^2a_5^{-\frac{11}{5}}
-\frac{8}{5}a_2a_4a_5^{-\frac{6}{5}}+a_1a_5^{-\frac{1}{5}},\cr
s=\frac{256}{3125}a_4^5a_5^{-4}-\frac{64}{125}a_3a_4^3a_5^{-3}
+\frac{16}{25}a_2a_4^2a_5^{-2}-\frac{4}{5}a_1a_4a_5^{-1}+a_0.
\end{eqnarray}\\
According to the cases 1-12 in section 4, it is easy to give the
classification of all traveling wave solutions to Eq.(\ref{KG}). We
omit them for briefly.

 \section{The solutions of irrational integrals and  applications}

 There are some mathematical physics equations whose reduced ODE's
 have the form of $(u')^2=\frac{g(u)}{f(u)}$, where $g(u)$ and $f(u)$ are
 two polynomials.
 So the classifications of their all traveling wave solutions
  are more complex.\\

\textbf{Example 1}.Classification of all single traveling wave
solutions
to Fujimoto-Watanabe equation\\

Fujimoto-Watanabe equation(\cite{FW}) reads
\begin{equation}
u_t=u^3u_{xxx}+3u^2u_xu_{xx}+3\varepsilon u^2u_x,
\end{equation}
under the traveling wave transformation $u=u(\xi), \xi=kx+\omega
 t$, the corresponding reduced ODE is as follows:
 \begin{equation}\label{FW6}
\pm(\xi-\xi_0)=\int\frac{u\mathrm{d}u}{\sqrt{a_3u^3+a_2u^2+a_1u+a_0}},
 \end{equation}
where $a_3=-\frac{2\varepsilon}{k}, a_1=-\frac{2\omega}{k^3}$,
$a_2$ and $a_0$ are two arbitrary constants. For simplicity, we
take the change of variable $ w=a_3^{\frac{1}{3}}u$, then the
Eq.(\ref{FW6}) becomes
 \begin{equation}\label{FW7}
\pm
a_3^{\frac{2}{3}}(\xi-\xi_0)=\int\frac{w\mathrm{d}w}{\sqrt{w^3+d_2w^2+d_1w+d_0}},
 \end{equation}
where $d_2=-a_3^{-\frac{2}{3}}a_2, \ \ d_1=a_3^{-\frac{1}{3}}a_1,
 \ \ d_0=a_0$. Denote $F(w)=w^3+d_2w^2+d_1w+d_0$, its complete
 discrimination system is
 $\triangle=-27(\frac{2d_2^3}{27}+d_0-\frac{d_1d_2}{3})^2-4(d_1-\frac{d_2^2}{3})^3$
 and $ D_1=d_1-\frac{d_2^2}{3}$. There are the following four
 cases to be discussed:\\

Case 1: $ \triangle=0, D_1<0 $, then we have $
F(w)=(w-\alpha)^2(w-\beta), \alpha\neq\beta. $ If $ w>\beta $, the
solutions are as follows:
\begin{equation}
\pm a_3^{\frac{2}{3}}(\xi-\xi_0)=2\sqrt{a_3^{\frac{1}{3}}u-\beta}+
\frac{\alpha}{\sqrt{\beta-\alpha}}\arctan\frac{\sqrt{a_3^{\frac{1}{3}}u-\beta}}{\sqrt{\beta-\alpha}},\
\ \alpha>\beta;
\end{equation}
\begin{equation}
\pm a_3^{\frac{2}{3}}(\xi-\xi_0)=2\sqrt{a_3^{\frac{1}{3}}u-\beta}+
\frac{\alpha}{2\sqrt{\alpha-\beta}}\ln|\frac{\sqrt{a_3^{\frac{1}{3}}u-\beta}-\sqrt{\alpha-\beta}}{\sqrt{a_3^{\frac{1}{3}}u-\beta}+\sqrt{\alpha-\beta}}|\
\ \alpha>\beta;
\end{equation}\\

Case 2: $ \triangle=0, \ D_1=0 $, then we have $ F(w)=(w-\alpha)^3
$, the solution is as follows:
\begin{equation}
\pm
a_3^{\frac{2}{3}}(\xi-\xi_0)=2\sqrt{a_3^{\frac{1}{3}}u-\beta}-\frac{\alpha}{\sqrt{a_3^{\frac{1}{3}}u-\beta}}.
\end{equation}

Case 3: $ \triangle>0, \ D_1<0 $, then $
F(w)=(w-\alpha)(w-\beta)(w-\gamma) $, we suppose that $
\alpha<\beta<\gamma $.  It is easy to see the corresponding
integral can be expressed by the second kind of elliptic
integrals.\\

Case 4: $ \triangle<0 $, then we have $ F(w)=(w-\alpha)(w^2+pw+q),
\ \ p^2-4q<0. $  Then the corresponding integral can be expressed
by the second kind of elliptic integrals.\\

\textbf{Example 2}. The classification of all single traveling wave
solutions to coupled Harry-Dym equation(\cite{CHD})
\begin{eqnarray}
u_t=\frac{1}{4}u_{xxx}-\frac{3}{2}uu_x+v_x+\varepsilon u_x,\\
v_t=-u_xv-\frac{1}{2}uv_x+\varepsilon v_x.
\end{eqnarray}
Through traveling wave transformation and integrations, we obtain
\begin{equation}
v=\frac{c_0}{(\frac{1}{2}ku+\omega-k\varepsilon)^2},
\end{equation}
\begin{equation}
(u')^2=\frac{2}{k^2}\frac{(u+r_0)^4+a_3(u+r_0)^3+a_2(u+r_0)^2+a_1(u+r_0)+a_0}{u+r_0},
\end{equation}
where $r_0=\frac{2\omega}{k}-2\varepsilon,
a_3=-4(\frac{\omega}{k}-\varepsilon, a_0=\frac{k^2}{2}c_1$. Denote
$F(u)=(u+r_0)^4+a_3(u+r_0)^3+a_2(u+r_0)^2+a_1(u+r_0)+a_0$,
according to the section 4, the complete discrimination system for
the fourth order polynomial $F(u)$ is as follows:
\begin{equation*}
\begin{split}
D_1=4, D_2=-p, D_3=8rp-2p^3-9q^2,\\
D_4=4p^4r-p^3q^2+36prq^2-32r^2p^2 -\frac{27}{4}q^4+64r^3,
\end{split}
\end{equation*}
\begin{equation}
F_2=9p^2-32pr,
\end{equation}
where $ p=a_2,\ \
 q=(\frac{a_3^3}{8}-\frac{a_2a_3}{2}+a_1), \ \
r=a_0-\frac{a_1a_3}{4}+\frac{a_2a_3^2}{16}-\frac{3a_3^4}{256}. $
There are the following cases to be discussed.\\

Case 1:$ D_4=0, D_3=0, D_2<0 $. Then we have
\begin{equation}
F(u)=(u+r_0)^2+l(u+r_0)+s^2,
\end{equation}
where $ l, s $ are real numbers, $ l^2-4s^2<0 $. We have
\begin{eqnarray*}
\pm(\xi-\xi_0)=\frac{1}{2\sqrt{2s-l}}\ln\frac{u+r_0\mp\sqrt{2s-l}\sqrt{u+r_0}+s}{u+r_0\pm\sqrt{2s-l}\sqrt{u+r_0}+s}
\end{eqnarray*}
\begin{eqnarray}
+\frac{1}{\sqrt{2s+l}}\{\arctan\frac{4\sqrt{u+r_0}\pm\sqrt{2s-l}}{2\sqrt{2s+l}}+\arctan\frac{4\sqrt{u+r_0}\mp\sqrt{2s-l}}{2\sqrt{2s+l}}\}.
\end{eqnarray}\\

Case 2: $ D_4=0, D_3=0, D_2=0 $. Then we have
\begin{equation}
F(u)=(u+r_0-\alpha)^4.
\end{equation}
 If $\alpha>0$,we have
\begin{eqnarray}
\pm(\xi-\xi_0)=\frac{1}{2\sqrt\alpha}\ln|\frac{\sqrt
{u+r_0}-\sqrt\alpha}{\sqrt {u+r_0}+\sqrt\alpha}|-\frac{\sqrt
{u+r_0}}{u+r_0-\alpha}.
\end{eqnarray}
If $\alpha<0$,we have
\begin{eqnarray}
\pm(\xi-\xi_0)=\frac{1}{\sqrt{-\alpha}}\arctan\frac{\sqrt
{u+r_0}}{\sqrt{-\alpha}}-\frac{\sqrt {u+r_0}}{u+r_0-\alpha}.
\end{eqnarray}\\

Case 3: $ D_4=0, D_3=0, D_2>0, E_2=0 $. Then we have
\begin{equation}
F(u)=(u+r_0-\alpha)^2(u+r_0-\beta)^2,
\end{equation}
where $ \alpha, \beta$ are real numbers, $ \alpha>\beta $. If
$\alpha>\beta>0$, we have
\begin{eqnarray}
\pm(\alpha+\beta)(\xi-\xi_0)=\sqrt\alpha\ln|\frac{\sqrt
{u+r_0}-\sqrt\alpha}{\sqrt
{u+r_0}+\sqrt\alpha}|-\sqrt\beta\ln|\frac{\sqrt
{u+r_0}-\sqrt\beta}{\sqrt {u+r_0}+\sqrt\beta}|.
\end{eqnarray}
If $0>\alpha>\beta$, we have
\begin{eqnarray}
\pm(\alpha+\beta)(\xi-\xi_0)=2\sqrt{-\alpha}\arctan\frac{\sqrt
{u+r_0}}{\sqrt{-\alpha}}-2\sqrt{-\beta}\arctan\frac{\sqrt
{u+r_0}}{\sqrt{-\beta}}.
\end{eqnarray}
If $\alpha>0>\beta$, we have
\begin{eqnarray}
\pm(\alpha+\beta)(\xi-\xi_0)=\sqrt\alpha\ln|\frac{\sqrt
{u+r_0}-\sqrt\alpha}{\sqrt
{u+r_0}+\sqrt\alpha}|-2\sqrt{-\beta}\arctan\frac{\sqrt
{u+r_0}}{\sqrt{-\beta}}.
\end{eqnarray}\\

Case 4. In all other cases such as
$F(u)=(u-\alpha)^2(u-\beta)(u-\gamma)$ and so on, the
corresponding solutions can be expressed with elliptic functions
and hyper-elliptic functions. We omit them for briefly.\\

\section{Classifications of solutions to some more general nonlinear differential equations}
There are many nonlinear differential equations which can be
reduced to an integrable ODE as follows
\begin{equation}\label{22}
u''(\xi)+p(u)(u'(\xi))^2+q(u)=0.
\end{equation}
These equations include High order KdV equation(\cite{HK})
\begin{equation}
u_t+u_x+\alpha uu_x+\beta
u_{xxx}+\alpha^2\rho_1u^2u_x+\alpha\beta(\rho_2uu_{xxx}+\rho_3u_xu_{xx})=0;
\end{equation}\\

Coupled KdV equation(\cite{CKDV})
\begin{eqnarray}
u_t+\varepsilon_1v_x+\varepsilon_2u^2u_x+\varepsilon_3u_{xxx}=0,\\
v_t+\eta_1(uv)_x+\eta_2vv_x=0;
\end{eqnarray}
Coupled mKdV equation(\cite{MKDV})
\begin{eqnarray}
u_t+\varepsilon_1vv_x+\varepsilon_2uu_x+\varepsilon_3u_{xxx}=0,\\
v_t+\eta_1(uv)_x+\eta_2vv_x=0,
\end{eqnarray}
and so on. We give the general solution of the Eq.(\ref{22}) in the following:\\

\textbf{Lemma}{\cite{lcs}}: The general solution of ODE (\ref{22})
is as follows:
\begin{equation}\label{ge}
\pm(\xi-\xi_0)=\int\frac{\mathrm{d}u}{\sqrt{\exp(-2\int
p(u)\mathrm{d}u)[c-2\int q(u)\exp(2\int p(u)
\mathrm{d}u)\mathrm{d}u]}},
\end{equation}
where $c$ and $\xi_0$ are two arbitrary constants.

We take the coupled KdV equation as an example to illustrate our
method for shortly. Other equations can be dealt with similarly.\\

\textbf{Example}: Classification of all single traveling wave
solutions to coupled KdV equation.

 Take the traveling wave transformation as $u=u(\xi), v=v(\xi), \ \ \xi=kx+\omega t
$, its reduced equations are as follows:
\begin{equation}\label{HD}
\gamma k^3u''+\frac{k\beta}{2}u^2-\frac{k\alpha}{2}v^2+\omega
u+c_0=0,
\end{equation}
\begin{equation}\label{uv}
u=-\frac{\delta}{2\rho}v-\frac{c_1}{\rho kv}-\frac{\omega}{\rho
k},
\end{equation}
where $c_0$ and $c_1$ are two integral constants. There are two
cases to be discussed.\\

Case 1. $c_1=0$. We have
$u=-\frac{\delta}{2\rho}v-\frac{\omega}{\rho k}$, and $v$
satisfies the following equation
\begin{equation}\label{3}
(v')^2=b_3v^3+b_2v^2+b_1v+b_0,
\end{equation}
where
\begin{equation}
b_3=\frac{2\rho}{\gamma k^2\delta}\{\frac{\alpha k}{3}+\frac{\beta
k\delta^2}{12\rho^2}\}, \ \
b_2=\frac{\beta\delta\omega-\delta\omega\rho}{\gamma
k^\delta\rho^2}, \ \ b_1=\frac{2\rho}{\gamma
k^2\delta}\{\frac{\beta
\omega^2}{k\rho^2}-\frac{2\omega^2}{k\rho}+2c_0\},
\end{equation}
and $b_0$ is an arbitrary constant. Denote $d_2=\frac{b_2}{b_3},
d_1=\frac{b_1}{b_3},d_0=\frac{b_0}{b_3}$
$F(v)=v^3+d_2v^2+d_1v+d_0$, and
$\triangle=-27d_0^2-4(-\frac{2d_1}{3})^3(-\frac{2d_2}{5})^{-1},
D_1=-\frac{2d_1}{3}(-\frac{2d_2}{5})^{-\frac{1}{3}}$. Here $
\triangle$ and $D_1$ make up a complete discrimination system for
$ F(v) $.We give  the classification of all solutions to the
Eq.(\ref{3}) as follows:\\

Case 1.1: If $ \triangle=0, D_1<0 $, then we have $
F(v)=(v-\alpha)^2(v-\beta), \alpha\neq\beta. $ If $ v>\beta $, the
solutions are as follows:
\begin{equation}
v=(\alpha-\beta)\tanh^2(\frac{b_3\sqrt{\alpha-\beta}}{2}(\xi-{\xi}_0))+\beta,\
\ \alpha>\beta;
\end{equation}
\begin{equation}
v=(\alpha-\beta)\coth^2(\frac{b_3\sqrt{\alpha-\beta}}{2}(\xi-{\xi}_0))+\beta,\
\ \alpha>\beta;
\end{equation}
\begin{equation}
v=(\beta-\alpha)\sec^2(\frac{b_3\sqrt{\beta-\alpha}}{2}(\xi-{\xi}_0))+\alpha,
\ \alpha<\beta.
\end{equation}\\

Case 1.2: If $ \triangle=0, \ D_1=0 $, then we have $
F(v)=(v-\alpha)^3 $, the solution is as follows:
\begin{equation}
v=4b_3(\xi-{\xi}_0)^{-2}+\alpha.
\end{equation}\\

Case 1.3: If $ \triangle>0, \ D_1<0 $, then $
F(v)=(v-\alpha)(v-\beta)(v-\gamma) $, we suppose that $
\alpha<\beta<\gamma $.  When $ \alpha<v<\beta $, we have
\begin{equation}
v=\alpha+(\beta-\alpha)\SN^2(\frac{b_3\sqrt{\gamma-\alpha}}{2}(\xi-{\xi}_0)),m);
\end{equation}\\
when $ v>\gamma $, we have
\begin{equation}
v=\frac{\gamma-\beta\SN^2(\frac{b_3\sqrt{\gamma-\alpha}}{2}(\xi-{\xi}_0)),m)}{\CN^2(\frac{b_3\sqrt{\gamma-\alpha}}{2}(\xi-{\xi}_0)),m)},
\end{equation}\\
where $ m^2=\frac{\beta-\alpha}{\gamma-\alpha} $.\\[3mm]

Case 1.4: If $ \triangle<0 $, then  $ F(v)=(v-\alpha)(v^2+pv+q), \
\ p^2-4q<0. $  We have
\begin{equation}
v=\alpha-\sqrt{\alpha^2+p\alpha+q}+\frac{2\sqrt{\alpha^2+p\alpha+q}}{1+\CN((\alpha^2+p\alpha+q)^{\frac{b_3}{4}}((\xi-{\xi}_0),m)}],
\end{equation}\\
where $ m^2=\frac{1}{2}(1-\frac{\alpha+\frac{p}{2}}{\sqrt{\alpha^2+p\alpha+q}}) $.\\

Case 2. $c_1\neq0$. According to the lemma, then $v$ satisfies the
following integral form:
\begin{equation}
\pm(\xi-\xi_0)=\int\frac{(v^2-\frac{2c_1}{k\delta})\mathrm{d}v}{v\sqrt{a_8v^8+\cdots+a_1v+a_0}},
\end{equation}
where $a_1, \cdots,a_8$ can be expressed by the corresponding
parameters. Denote $F(v)=a_8v^8+\cdots+a_1v+a_0$, its complete
discrimination system is too complex to write. We omit these
solutions for briefly.

\section{Discussions and conclusions  }

In Ref. \cite{FAN3}, Fan proposed an expanse method to seek for the
single traveling wave solutions to nonlinear partial differential
equations. His key idea is to assume that the solution has a
polynomial form $u=a_n\phi^n+\cdot+a_1\phi+a_0$ of a function $\phi$
which satisfies an ODE as follows:
\begin{equation}
\phi'=\pm\sqrt{c_r\phi^r+\cdot+c_1\phi+c_0},
\end{equation}
According to the balance principle, a relation of $n$ and $r$ can
be obtained. In his paper, Fan take a special case of $r=4$ as an
example to apply his method. In the end of his paper, he said "we
only have investigated a special case when $r=4$. the proposed
method can be extended to the cases when $r>4$. The details for
those cases will be investigated in our future works". In fact,
there are two problems need to answer. One problem is whether or
no there exist new solutions in the cases of $r>4$ for some
equations. The second problem is how to prove some solutions are
new in the cases of $r>4$ if the equation considered can't be
directly reduced to an integral form.

According to the above analysis to direct integral method ,we can
make a conclusion that there aren't new solutions in the cases of
$r>4$ for those equations which can be directly reduced to an
integral form. Thus we have answered the Fan's first
problem.\\

Direct integral method is very natural and simple for solving
traveling wave solutions to nonlinear partial differential
equations. But it is just complete discrimination system for
polynomial make direct integral method becomes an efficient and
powerful method. A lot of nonlinear mathematical physics equations
can be dealt with by direct integral method, their all traveling
wave solutions can be classified. These complete results can't be
obtained by using of other methods. On the other hand,  a lot of
nonlinear evolution equations such as (1+1)-dimensional Boussinesq
equation and (3+1)-dimensional KP equation and so on, can be reduced
to integral forms by setting some integral constants to zeroes. For
those equations, we can not give their all traveling wave solutions,
but we can still obtain their abundant exact solutions. For example,
we consider the (2+1)-dimensional boussinesq equation
\begin{equation}\label{a1}
u_{tt}-u_{xx}+3(u^2)_{xx}-u_{xxxx}-u_{yy}=0.
\end{equation}
Instituting traveling wave transformation $ u=u(\xi),
\xi=k_1x+k_2y+\omega t$ into Eq.(\ref{a1}) and integrating twice
yield
\begin{equation}\label{a2}
k_1^4u''=(\omega^2-k_1^2-k_2^2)u+3k_1^2u^2+c_2\xi+c_1.
\end{equation}
where $c_1$ and $c_2$ are two arbitrary constants. If we take
$c_2=0$, then we can apply the direct integral method to it. In
fact, integrating Eq.(\ref{a2}) once yields
\begin{equation}\label{a3}
\pm k_1^2(\xi-\xi_0)=\int
\frac{\mathrm{d}u}{\sqrt{k_1^2u^3+(\omega^2-k_1^2-k_2^2)u^2+c_1u+c_0}},
\end{equation}
where $c_0$ is an arbitrary constant. According to Ref.[19], we can
easily to give classification of all solutions to Eq.(\ref{a3}).
Thus it is easy to obtain abundant traveling  wave solutions to
Eq.(\ref{a1}). For briefly, we omit them.

\end{document}